\documentclass[sigconf]{acmart}

\AtBeginDocument{%
  }

\setcopyright{acmlicensed}
\copyrightyear{2018}
\acmYear{2018}
\acmDOI{XXXXXXX.XXXXXXX}

\acmConference[MSR 2024]{MSR '24: Proceedings of the 21st International Conference on Mining Software Repositories}{April 15–16, 2024}{Lisbon, Portugal}

\acmISBN{978-1-4503-XXXX-X/18/06}

\usepackage{url}
\usepackage{multirow}

\usepackage{listings}

\usepackage{xcolor}
\usepackage{graphicx}
\usepackage{mdframed}
\usepackage{enumitem}
\colorlet{punct}{red!60!black}
\definecolor{background}{HTML}{EEEEEE}
\definecolor{delim}{RGB}{20,105,176}
\definecolor{darkblue}{rgb}{0.0, 0.0, 0.5}
\colorlet{numb}{magenta!60!black}

\usepackage[strict]{changepage}
\usepackage{framed}

\definecolor{formalshade}{rgb}{0.95,0.95,1}

\newenvironment{quotebox}{%
  \MakeFramed{\advance\hsize-\width\FrameRestore}%
  \noindent\hspace{-4.55pt}%
  \begin{adjustwidth}{}{7pt}%
  \vspace{2pt}\vspace{2pt}%
}
{%
  \vspace{2pt}\end{adjustwidth}\endMakeFramed%
}
\usepackage{hyperref}

\begin{document}

\title{The role of library versions in Developer-ChatGPT conversations}

\author{Rachna Raj, Diego Elias Costa}
\affiliation{%
  \institution{Department of Computer Science and Software Engineering\\ Concordia University}
  \streetaddress{P.O. Box 1212}
  \city{Montreal}
  \state{Quebec}
  \country{Canada}
  \postcode{43017-6221}
}
\email{rachna.raj@mail.concordia.ca, diego.costa@concordia.ca}

\renewcommand{\shortauthors}{Trovato et al.}

\begin{abstract}

The latest breakthroughs in large language models (LLM) have empowered software development tools, such as ChatGPT, to aid developers in complex tasks. 
Developers use ChatGPT to write code, review code changes, and even debug their programs.
In these interactions, ChatGPT often recommends code snippets that depend on external libraries. 
However, code from libraries changes over time, invalidating a once-correct code snippet and making it difficult to reuse recommended code. 

In this study, we analyze DevGPT, a dataset of more than 4,000 Developer-ChatGPT interactions, to understand the role of library versions in code-related conversations. We quantify how often library version constraints are mentioned in code-related conversations and when ChatGPT recommends the installation of specific libraries. Our findings show that, albeit to constantly recommend and analyze code with external dependencies, library version constraints only appear in 9\% of the conversations. In the majority of conversations, the version constraints are prompted by users (as opposed to being specified by ChatGPT) as a method for receiving better quality responses. Moreover, we study how library version constraints are used in the conversation through qualitative methods, identifying several potential problems that warrant further research.

\end{abstract}

\maketitle

\section{Introduction}

In the field of Artificial Intelligence \cite{hou2023large}, large language models (LLM) like OpenAI's ChatGPT and Google Gemini, are promising to revolutionize our technology.  
Trained on massive data, LLM models like ChatGPT have pushed the boundaries across multiple application areas, and have shown remarkable advances in software development, from code summarization~\cite{CodeSummarizationWithLLMs} and understanding~\cite{CodeUnderstandingWithLLMs}, to program synthesis~\cite{ProgramSynthesisWithLLMs} and automated program repair~\cite{RepairWithLLMs}.
Given its capacity to understand and carry out complex contexts across multiple interactions, ChatGPT has been deemed to provide equal or better assistance than community-based forums like Stack Overflow~\cite{betterProgAssist}.

At the core of ChatGPT's remarkable performance in helping software developers lies in its capability to understand and recommend source code. 
Modern code is heavily dependent on open-source libraries~\cite{Zajdel:OSS}, consequently, ChatGPT is constantly bombarded with prompts that involve recommending code that uses APIs from external dependencies.
As libraries evolve, library developers constantly change their APIs to provide better functionality, cater to new users, or fix design problems, leading to breaking changes that invalidate a once-correct code.~\cite{Venturini:23:BreakingChange}.
Breaking changes have been at the center of dependency problems, leading to bad dependency practices~\cite{Jafari:22:Smells}, risks of vulnerabilities~\cite{Alfadel_Costa_Shihab_2023}, and technical lags~\cite{Decan:SemanticVersioning}.
Research suggests that ChatGPT can provide quality code suggestions in library-related tasks~ \cite{betterProgAssist}, but is ChatGPT aware of the differences across library versions and their consequences for the correctness of code snippets?

In this study, we aim to conduct the first preliminary study on ChatGPT awareness of library versions.  
We explored a dataset\cite{DevGPT} of Developer-ChatGPT conversations to understand the role of library versions in the conversation between users and ChatGPT, when discussing code with external dependencies. 
To understand the role of library versions, we employ a mixed-method approach, combining both quantitative and qualitative analyses, on a dataset with more than 4,000 conversations between developers and ChatGPT.
We formulate our study to answer two main research questions:

\textbf{RQ1. How often are library versions discussed in library-related conversations?}
We find that the vast majority (88.5\%) of conversations that include libraries in code snippets make no mention of library versions. This lack of version discussion is true even when ChatGPT recommends the installation of a new dependency, which implicitly indicates a bias towards the dependencies' latest version.

\textbf{RQ2. What is the role of library versions in library-related conversations?}
Library versions are frequently used to fine-tune ChatGPT responses and are often specified by developers by their own dependency configuration files in their prompts. There are also some attempts at using ChatGPT to resolve incompatibility issues caused by ChatGPT's own responses, to varying results.

\section{Dataset}

\label{sec:dataset}

This study aims to understand the role of library versions in developer-ChatGPT interactions. 
To that aim, we mine the DevGPT \cite{DevGPT} dataset, a dataset of Developer-ChatGPT conversations with more than 2,000 conversations~\cite{xiao2023devgpt}.  
DevGPT \cite{DevGPT} was created by Xiao et al.~\cite{xiao2023devgpt} by mining links to ChatGPT conversations shared by developers using the OpenAI sharing functionality~\footnote{\url{https://help.openai.com/en/articles/7925741-chatgpt-shared-links-faq}}
The dataset is a collection of JSON files covering the interaction metadata, prompts asked and answers from ChatGPT, including text and code snippets.

As the DevGPT \cite{DevGPT} is collected by mining open-source sources, such as GitHub and Hacker News, the authors publish a new snapshot periodically, with the first version published on July 27, 2023.
We opted to select the latest version available for our study (at the time of analysis), which was the 9th snapshot, published on October 24, 2023.
This snapshot included a total of {4116} Developer-ChatGPT conversations,

As we are interested in understanding the context and the code-related content of Developer-ChatGPT conversations, we explore primarily the URL links provided, the metadata for filtering target conversations, and the content exported within the JSON field \texttt{ChatGPTSharing}. 
The field \texttt{ChatGPTSharing} contains metadata like the URL for that conversation, and it also includes a list of conversation objects storing many single \texttt{Conversation} instances. 
For the purposes of our study, we name the individual conversation instances as \textit{interactions} as they are composed of a single \texttt{Prompt}, a single \texttt{Answer}, and a \texttt{ListOfCode} field that structures the code snippets and their respective programming language.

The DevGPT\cite{DevGPT} dataset contains Developer-ChatGPT conversations coming from six distinct sources:
\begin{itemize}

    \item \textbf{Pull Request} includes conversations from GitHub pull requests where developers request ChatGPT's assistance in code-related questions and problem-solving.
    
    \item \textbf{Commits} includes conversations that were shared within a repository commit. 
    Commits are usually more fine-grained than pull requests, and offer the opportunity for analysing a more targeted conversation between developers and ChatGPT.

    \item \textbf{Issue} include conversations that were shared in GitHub issues. 
    These conversations often target troubleshooting software development problems, such as fixing a bug or an environment configuration.  
    
    \item \textbf{File sharing} includes conversations that were included in source-code files from GitHub.

\end{itemize}
    
We exclude Hacker News and Discussion as sources for our study because they don't directly apply ChatGPT recommendations to specific software projects.

\begin{table*}
    \caption{Statistics on the DevGPT dataset filtered by our study. Our targeted languages for this study are JavaScript, TypeScript, Python, and Java.}
    \label{tab:dataset_stats}
    \centering
    \begin{tabular}{l r r r r|r}
    \toprule
         \textbf{Dataset} & \textbf{Pull Request} & \textbf{File sharing} & \textbf{Commits} & \textbf{Issues} & \textbf{Total} \\
    \midrule
       All conversations (original dataset) &286 &2540 & 692 & 598 & 4116 \\
        Code-related conversations &179 &1184 &674 &360 & 2397 \\
        Code-related conversations from targeted languages &100 &614 &63 &196 & 973 \\
        Library-related conversations & 53 & 286&40& 107 & 486 \\
        Conversations with download suggestion& 15 & 86 &7& 23 & 131 \\
   
    \bottomrule
    \end{tabular}

\end{table*}

\section{Methodology}

The goal of our study is to perform a preliminary analysis of the role of library versions in Developer-ChatGPT conversations. 
The DevGPT\cite{DevGPT} dataset has numerous types of conversations related to software development, including ones that are unrelated to code or do not contain external dependencies. 
We need to identify \textbf{library-related conversations}.
Library-related conversations are conversations that 1) contain code snippets and 2) include external libraries in the code snippets. 
In the following, we detail how we identify these target conversations to answer two core research questions of our study: 

\begin{itemize}
    \item \textbf{RQ1}: How often do versions appear in library-related conversations?

    \item \textbf{RQ2}: What is the role of library version in library-related conversations?
\end{itemize}

\subsection{Finding code-related conversations.}
\label{sub:finding-code-related-conversations}

The DevGPT\cite{DevGPT} dataset groups all code snippets given by ChatGPT in its answer in a specific JSON field called  \texttt{ListOfCode.} 
The ListOfCode object has two relevant fields for our study:
the \texttt{Type} which specifies the programming language and the \texttt{Content} which includes the code snippet itself.
We consider a \textbf{code-related conversation}, a conversation that contains a code snippet in one or more of its interactions (prompt from user and answer from ChatGPT).
Thus, we filter conversations that contained a valid \texttt{ListOfCode} object, with a code snippet in its content field. 
From the 4116 conversations in the DevGPT dataset, we find that \textbf{2397 (58.2\%) are code-related conversations}, i.e., conversations are linked with at least one code snippet.

To make our study more manageable, we decided to focus our analysis on conversations targeting four major programming languages:  
Java, Python, JavaScript and TypeScript. 
These four programming languages are very popular among software developers~\cite{StackOve7:online}, are widely represented in the DevGPT dataset~\cite{xiao2023devgpt}, and include large ecosystems of reusable software dependencies~\cite{Decan_Mens_Grosjean_2019, Mujahid:NPM}.
Maven, PyPI and NPM are among the largest software ecosystems to date, with each containing between 500 thousand to 4 million reusable packages~\cite{TIDELIFT:online}, a fertile ground for researching software dependency practices. 
To filter conversations that included Java, JavaScript, TypeScript, and Python code, we filter conversations where the field \texttt{Type} points to our target programming languages.
Table~\ref{tab:dataset_stats} shows that from the 2397 code-related conversations, \textbf{973 (40.5\%) are code-related conversations about Java, Python, JavaScript and TypeScript code.}
While targeting four specific languages reduced our dataset size, focusing on specific languages allowed us to employ more precise methods for identifying libraries in code and their respective versions, as these ecosystems employ similar versioning strategies~\cite{Semantic15:online}.

\subsection{Filtering library-related conversations.}

From the code-related conversations, we now need to identify the ones that include external dependencies in their code snippets, i.e., the library-related conversations.
We consider library-related conversation, any conversation where external libraries are imported in the code snippet, either from the developer or from ChatGPT. 
Since we focused on code-related conversations from Java, Python, and Javascript, we resort to identifying when libraries are imported in those languages.  
These languages allow developers to import libraries by using the special keyword \texttt{import} in the code. 
We also considered the keyword \texttt{require} for JavaScript related code, a more versatile method for importing libraries. 
We search for these two keywords in the code snippet of all conversations, and include a conversation if one or more code snippets included any dependency. 
As Table~\ref{tab:dataset_stats} shows, from the 973 code-related conversations from our four target languages, \textbf{we find that 486 (49.9\%) conversations included external dependencies} in at least one code snippet, i.e., are library-related conversations.

\textbf{Finding library download suggestions.}
Note that, while we coined conversations as "library-related", the conversations are not necessarily primarily about libraries.
However, We expect library versions to play a greater role when ChatGPT explicitly mentions a specific library to be installed.
To test this hypothesis, we filtered ChatGPT conversations that explicitly recommend users to install libraries, by searching for commands such as: ``pip install'', ``conda install'', ``npm install'', or ``mvn install'', as these are the de facto methods for installing libraries in the targeted programming languages.
We show in Table~\ref{tab:dataset_stats} that from 486 library-related conversations, \textbf{129 (26.5\%) have explicit mentions of installing new dependencies}.

\vspace{-2mm}
\subsection{Finding library versions in library-related conversations}

We start our analysis with 486 library-related conversations, with each conversation containing multiple interactions (rounds of prompts and answers).  
Understanding the role of library version in these conversations, requires us to identify library versions from the potentially large text from these conversations.  
To that aim, we rely on the practices of semantic versioning~\cite{Decan:SemanticVersioning}.
In all our targeted programming languages, library developers are encouraged to version their libraries using a \textbf{\texttt{major}.\texttt{minor}.\texttt{patch}} format. 
In fact, not using or following semantic versioning is perceived as a bad practice in these software ecosystems, affecting the popularity and engagement from the library users~\cite{Jafari:22:Smells}.

To capture conversations that include at least a single mention of library versions, we employ a regex that captures semantic versioning specifications~\cite{Semantic15:online}. 
We consider relevant to our study, any conversations that include library versions either in the user prompts or in the ChatGPT responses (including in the code snippets).
Our method of using regex to capture versions using semantic specifications is prone to false-positives.
We manually analyze the conversations and exclude false-positives, usually related to IP code that follows similar textual structure.

\subsection{Qualitative analysis of version-related conversations}

We use a qualitative approach to identify the role of library versions in Developer-ChatGPT conversations. 
For each conversation, the two authors have jointly discussed the following aspects:
\begin{itemize}

    \item \textbf{Who} first specifies the library version in the conversations? 
    We classify conversations as \textbf{proactive}, if ChatGPT, when recommending code, describes for which library version ranges that code is valid. 
    If a version is mentioned by the user as a response to a ChatGPT answer, we classify the conversation as \textbf{reactive}.
    
    \item \textbf{What} is the role of library version in the conversation? Library version may be specified by the user when troubleshooting incompatibilities or dependency conflicts, asking for specific API versions or library recommendations, etc.    

\end{itemize}

For this question, we follow an inductive approach, where the codes from the above subquestions were developed while reading the conversations. 
We used an open coding method where the codes emerged directly from the contents under analysis. \cite{Saldana:2009}. 
The authors opted for jointly coding the conversations, discussing eventual disagreements and reaching consensus.  
As part of this process, we also identified and excluded 3 library-related conversations that had non-English text, as well as 5 library-related conversations that had no external dependence. Usually, these conversations involve apps, utilities, or internally built libraries.

\section{Findings}

\begin{table}
    \caption{How often do dependency-related conversations include mentions of library versions?}
    \vspace{-1mm}
    \label{tab:rq1_results}
    \centering
    \begin{tabular}{l l r r}
        \toprule
        &  & Total & Version \\
        \midrule
      \multirow{5}{*}{Conversations} & Pull Request & 53 & 8  \\
        & File Sharing & 286 & 20  \\
        & Commits &40 & 11 \\ 
        & Issues &107 & 8  \\
        & \textbf{Total} & \textbf{486}    & \textbf{47 (9.67\%)}          \\ 
        \midrule
       \multirow{5}{*}{Download suggestions} & Pull Request & 15 & 4 \\
        & File Sharing &86 &10\\
        & Commits &7 & 3 \\
        & Issues &23 & 7 \\
        & \textbf{Total} & \textbf{131}    & \textbf{24 (18.3\%)}          \\ 
         \bottomrule
    \end{tabular}

\end{table}
\subsection{How often are library versions discussed in library-related conversations?}
\label{sub:rq1-findings}

\noindent

We report the results of this research question in table~\ref{tab:rq1_results}.
We observe that, when considering all dependency-related conversations, \textbf{only 47 out of 486 (9.67\%) have mentions of library versions}.
The proportion remains similar across different sources, with commits, being the category with a higher frequency of versions being part of the conversation. 

Looking at the conversations that explicitly recommend the installation of dependencies (the bottom part of Table~\ref{tab:rq1_results}), it is surprising to see that library versions are similarly frequent. 
We note that only in 24 out of 131 conversations (18.2\%), versions of libraries are explicitly mentioned. 
This result has an important implication: when recommending users dependency-related code or specific library installations, ChatGPT implicitly recommends the latest library version.

\begin{quotebox}
    \textbf{Discussion:} The majority of dependency-related conversations have no mentions of library versions, even when ChatGPT explicitly recommends users to install new libraries. 
    \\
    \textbf{Question for future research:} Is ChatGPT code suggestions often compatible with the latest version of libraries? How do recent breaking changes affect the accuracy of ChatGPT code recommendations? 
\end{quotebox}

\subsection{What is the role of library versions in library-related conversations?}
\label{sub:rq2-findings}

Table~\ref{tab:rq2-findings} presents the overall results of our qualitative analysis. Given the small dataset, we refrain from presenting percentages and focus only on the patterns that emerge across at least two distinct conversations. Note that a conversation could be categorized across multiple roles, as a single conversation includes multiple interactions between developers and ChatGPT. 

\textbf{Who first specifies the library version?}
The first insight that emerges is that, in the vast majority of the analyzed conversations, 39 out of 47, the user provides ChatGPT with library versions either in the first prompt or in response to ChatGPT's responses.
After the library version is mentioned, ChatGPT carries out the mentioned range in future responses. 
In the other 10 conversations, ChatGPT proactively mentions the library version when guiding users through a setup process or suggesting unit test code.
In this pull request~\cite{ChatGPT:proactive}, ChatGPT provides the version of the testing library when instructing the user on how to test their application.

\textbf{What is the role? Configuration setup.}
A major role of library versions in Developer-ChatGPT conversations is to help prune ChatGPT responses. 
We noticed a pattern that developers frequently provide their dependency configuration file as part of the prompt to ChatGPT, to help ChatGPT understand the dependencies their project has. 
For example, in the PR~\cite{Refactor7:online}, the author asks ChatGPT to help them in refactor their project code, and include amongst all their source code and package.json file, to enrich the context of the query.

\textbf{Troubleshooting incompatibility issues.}
The second most common role library versions have is related to troubleshooting dependency conflicts. 
We found this case in 10 conversations, and in all but one case, the dependency conflict was caused by ChatGPT responses. 
For example, in this long conversation~\cite{Restruct22:online}, the developers frequently paste error logs related to incompatibilities that include library version ranges, prompting ChatGPT to recommend solutions.

\textbf{Pinning Dependencies.}
Eight conversations showed a specific configuration setup, where developers include a specific version of a library passed as a parameter within the code. 
For example, in this conversation~\cite{Restruct23:online}, a developer shared code with JQuery version parameter, seeking help with an error in node.js. ChatGPT explained the missing feature and suggested an alternative library, but failed to resolve the issue.

\begin{quotebox}
    \textbf{Discussion:} Developers are the ones that more frequently specify library versions in conversation with ChatGPT, with many users appending their own dependency configuration file to their prompt in the hopes of getting tailor-made responses for their dependencies. 
    \\
    \textbf{Question for future research:} How effectively can ChatGPT map library version ranges and their respective valid APIs? Can ChatGPT be used as a trustworthy source of information for finding incompatibility problems? 
\end{quotebox}

\begin{table}
    \caption{What is the role of library version in library-related conversations? The frequency reported is out of 47 coded conversations.}
    \label{tab:rq2-findings}
    \centering
    \begin{tabular}{l l r}
    \toprule
        \textbf{Category} &
        \textbf{Qualitative analysis} & \textbf{Freq.} \\
        \midrule
       \multirow{2}{*}{Who?} & ChatGPT reacts to the requested version  & 39 \\
        &ChatGPT proactively recommends versions & 8 \\

        \midrule
        \multirow{3}{*}{What?} &Configuration setup & 13 \\ 
        & Troubleshoot incompatibility issues & 10 \\
        &  Pinning Dependencies & 8\\ 
        &  Others & 28\\ 

    \bottomrule
    \end{tabular}

\end{table}

\vspace{-2mm}
\section{Threats to Validity}

There are a few important limitations to our work that need to be considered when interpreting our findings.
First, DevGPT dataset is reasonably small and may include biases that would prevent it from being representative of the usual developer-ChatGPT interactions. 
Second, while we verify the precision of our process for identifying library versions in conversations, our method may still miss relevant conversations. Users may communicate more casually to ChatGPT, and phrases such as "Is this code compatible with versions greater than 4.x.x of library X?", would not be captured by our method. 
As such, our quantitative results should be interpreted as a lower bound number for conversations with mentions of library versions.

\section{Conclusion}
This paper presents a preliminary study aiming to grasp an understanding of the role of library versions in Developers-ChatGPT interactions. 
Our study shows that library versions are only rarely part of the vocabulary of Developers-ChatGPT conversations (RQ1). Users employ different strategies to trigger a more library version-aware responses from ChatGPT to varying results (RQ2). Finally, we point out some questions that may spark further research in the topic.

\section{Data Availability}

The source code and dataset created as a part of this research are available at \url{https://github.com/rachnaraj/MSR-RR_Mining_Challenge2023}.

\bibliographystyle{ACM-Reference-Format}
\bibliography{bibliography}

\end{document}